\documentclass[sn-basic,onecolumn]{sn-jnl}

\usepackage{mathtools}
\usepackage{amsmath,amssymb,amsfonts,bm}
\usepackage{graphicx}
\graphicspath{{figs/}}

\usepackage{enumitem}
\usepackage{subcaption}
\usepackage{float,placeins}

\definecolor{pumpkin}{rgb}{1.0, 0.46, 0.09}

\usepackage{chngcntr}
\counterwithin{figure}{section}
\usepackage{longtable} 
\usepackage{dblfloatfix}
\usepackage{lineno}
\usepackage{setspace}
\usepackage{hyperref}
\hypersetup{
    colorlinks=true,
    linkcolor=blue,
    filecolor=magenta,      
    urlcolor=blue,
}
\usepackage[para,online,flushleft]{threeparttable}
\usepackage{multirow}

\jyear{2022}%

\theoremstyle{thmstyleone}%
\theoremstyle{thmstyletwo}%

\theoremstyle{thmstylethree}%

\usepackage{csquotes}
\usepackage{siunitx}

\begin{document}

\title[Bayesian Assessment of Heavy Metal Health Risk]{Quantifying the Threat: A Bayesian Probabilistic Framework for Heavy Metal Health Risks among Children in the Densu Basin, Ghana} 


\author*[1]{\fnm{T.} \sur{Ansah-Narh}}\email{theophilus.ansah-narh@gaec.gov.gh}    
\author[1]{\fnm{G. Y.} \sur{Afrifa}}
\author[1]{\fnm{J. B.} \sur{Tandoh}}
\author[2]{\fnm{E. N. N.} \sur{Nortey}} 
\author[3]{\fnm{D.} \sur{Adjei}} 
\author[4]{\fnm{R.} \sur{Opoku-Sarkordie}}
\author[5]{\fnm{R.} \sur{Davis}}






 \affil[1]{\orgdiv{Ghana Space Science and Technology Institute}, \orgname{Ghana Atomic Energy Commission}, \orgaddress{ \postcode{P. O. Box LG 80}, \state{Accra}, \country{Ghana}}}
 \affil[2]{\orgdiv{Department of Statistics and Actuarial Science}, \orgname{University of Ghana},  \postcode{P. O. Box LG 115}, \orgaddress{\street{Legon}, \city{Accra}, \country{Ghana}}}
\affil[3]{\orgdiv{Institut de Chimie Physique}, \orgname{Université Paris-Saclay / CNRS UMR8000},  \postcode{91400}, \orgaddress{\street{Rue Michel Magat}, \city{Orsay}, \country{France}}}
\affil[4]{\orgdiv{Department of Information Technology and Mathematical Sciences}, \orgname{Methodist University Ghana},  \postcode{P. O. Box DC 940}, \orgaddress{\street{Dansoman}, \city{Accra}, \country{Ghana}}}

 \affil[5]{\orgdiv{Department of Mathematics and Actuarial Science}, \orgname{Pentecost University},  \postcode{P. O. Box KN 1739}, \orgaddress{\street{Kaneshie}, \city{Accra}, \country{Ghana}}}





\abstract{
Groundwater remains a critical source of domestic water for many communities in the Densu River Basin, Ghana, yet contamination by potentially toxic elements poses persistent public health concerns, particularly for children. Existing health risk assessments in the region have largely relied on deterministic hazard index calculations, which offer limited insight into uncertainty, co-exposure effects, and the spatial variability of risk. This limitation constrains evidence-based decision-making in hydrogeologically complex and data-limited settings.
To address this gap, this study develops and applies a Bayesian probabilistic health risk assessment framework to quantify non-carcinogenic risks associated with multi-metal groundwater exposure among children. Concentrations of Fe, Mn, Ni, Pb, Cd, and As, measured in $94$ groundwater samples, were integrated into a Bayesian regression model that explicitly propagates parameter uncertainty and jointly accounts for contaminant co-occurrence. Posterior inference was performed using adaptive Hamiltonian Monte Carlo sampling, enabling robust estimation of metal-specific risk contributions and posterior predictive health outcomes. Site-level posterior risk probabilities were subsequently translated into spatially continuous risk surfaces using inverse distance weighting, ensuring consistency with sampling density and hydrogeological constraints.
The results indicate that lead, arsenic, and cadmium are the dominant contributors to cumulative health risk, with posterior probabilities providing strong evidence of their positive association with the child Health Index. Posterior predictive analysis reveals that approximately $31\%$ of sampling locations exhibit probabilities exceeding $0.5$ for $\mathrm{HI} > 1$, indicating that unsafe exposure is more likely than not at these sites under the assumed exposure conditions. The spatial risk surface is characterised by substantial heterogeneity, with a posterior mean risk probability of $0.372$ and a standard deviation of $0.416$, reflecting both pronounced variability in groundwater chemistry and uncertainty propagated through the Bayesian model. Importantly, high-risk zones emerge that are not readily identifiable from individual metal concentration maps alone, demonstrating that cumulative risk often arises from moderate co-exposure rather than extreme contamination by a single metal.
By moving beyond binary threshold-based classification, the proposed framework provides uncertainty-aware, spatially explicit risk information that is directly relevant for groundwater management and public health planning. The approach offers a transparent and adaptable tool for prioritising monitoring, mitigation, and policy interventions in groundwater-dependent regions, with broader applicability across sub-Saharan Africa.
}
 
\keywords{Densu river basin, Heavy metals, Groundwater contamination, Bayesian health risk assessment, Spatial risk mapping, Hydrogeochemical uncertainty}

\maketitle
\section{Background and Motivation}
\label{sec:introduction}

Groundwater constitutes a cornerstone of domestic, agricultural, and industrial water supply worldwide, particularly in regions where surface water availability is constrained by climatic variability, seasonal intermittency, and infrastructural limitations. It is estimated that nearly half of the global population relies on groundwater for drinking and daily use, with substantially higher dependence observed in rural and peri-urban areas of low- and middle-income countries \citep{world2024guidelines}. This reliance places considerable importance on groundwater quality, especially in settings where alternative water sources and advanced treatment options are limited.

Despite its widespread perception as a relatively protected resource, groundwater quality is increasingly compromised by both natural and anthropogenic processes. Geochemical weathering of aquifer materials can mobilise trace elements, while agricultural intensification, industrial activities, mining, and improper waste disposal contribute to potential contamination pathways. Potentially toxic elements (PTEs), including lead (Pb), cadmium (Cd), and arsenic (As), are of particular concern because of their persistence in subsurface environments and their potential to cause adverse health effects following prolonged exposure \citep{kapoor2024health,Smedley2002,raja2021health}. Chronic ingestion of contaminated groundwater has been associated with a range of adverse health outcomes, underscoring the need for robust and context-specific risk assessment frameworks.

The occurrence and distribution of PTEs in groundwater depend on the interaction between geological characteristics, hydrogeochemical conditions, groundwater flow pathways, and land-use pressures. Weathering of mineral-bearing formations may release elements into groundwater, whereas changes in redox conditions, pH, and mineral-surface interactions can alter their mobility and persistence. Anthropogenic activities may introduce additional contaminants or modify existing geochemical conditions, resulting in spatially heterogeneous contamination patterns. Research on lead-zinc mine tailings in South China has examined the spatial distribution of heavy metals and their environmental implications, illustrating the relevance of spatially explicit characterisation in contaminated environments \citep{Chen2018}. More recently, Isah et al. investigated heavy metal contamination associated with artisanal gold mining in the Ile-Ife--Ilesha Schist Belt of Nigeria, using geospatial assessment to examine pollution sources and contamination hotspots \citep{Isah2025a}. Although mining and mine-tailings environments differ from the groundwater-dependent setting examined in this study, these investigations demonstrate the importance of considering spatial heterogeneity and potential contamination sources when characterising environmental metal pollution.

Children represent a population group that is especially susceptible to the health impacts of PTE exposure through drinking water. Relative to adults, children typically consume greater volumes of water per unit body weight and undergo critical stages of neurological, immunological, and physiological development. These characteristics can increase exposure on a body-weight basis and heighten vulnerability to certain toxic substances \citep{zhao2023health,hui2021cumulative}. Epidemiological studies have consistently linked early-life exposure to metals such as Pb and Cd with impaired neurodevelopment, diminished cognitive function, and elevated risks of long-term morbidity \citep{grandjean2019developmental,lanphear2019erratum,choi2012developmental}. Consequently, international public health guidelines emphasise the need to evaluate exposure in vulnerable population groups, including children in groundwater-dependent communities. For these populations, assessments that consider the combined contribution of multiple contaminants are particularly relevant because exposure may occur simultaneously through water containing several potentially harmful elements.

The challenges associated with groundwater contamination are particularly pronounced in sub-Saharan Africa, where groundwater serves as an important source of potable water. Rapid population growth, unplanned urbanisation, agricultural expansion, extractive activities, and inadequate waste management can intensify pressures on groundwater systems across the region. Previous assessments have documented contamination associated with geogenic processes and anthropogenic inputs, including mining and waste disposal \citep{owusu2024water,lapworth2017urban}. In southwestern Nigeria, Isah et al. investigated groundwater-level conditions and heavy metal contamination around the Olubunku dumpsite and its surroundings \citep{Isah2023}. Their subsequent work examined groundwater contamination flow paths and heavy metal mobilisation near a waste site in the same broader region \citep{Isah2025b}. These studies are relevant to the present investigation because they address the relationship between groundwater conditions, contaminant mobilisation, and potential sources of pollution. However, the transferability of their findings to other aquifer systems must be considered in relation to differences in geology, hydrogeology, land use, and contaminant sources.

Despite growing attention to groundwater contamination, monitoring remains spatially sparse and temporally irregular in many settings, limiting the ability to characterise contaminant distributions and associated exposure patterns comprehensively. Concentration measurements from individual sampling locations provide important information about local water quality, but they do not, in isolation, establish the probability of adverse health-risk thresholds being exceeded across a population or geographical area. Spatial variability in contaminant concentrations also means that basin-wide summaries may conceal localised areas of concern. These challenges are particularly important where groundwater is used directly for drinking and where resources for routine monitoring and treatment are limited.

Within this broader regional context, Ghana exemplifies a setting where groundwater plays a critical role in meeting domestic and productive water demands. The Densu Basin supports a dense and rapidly growing population and encompasses a heterogeneous mosaic of land uses, including urban settlements, agriculture, and industrial activities. Previous studies have documented spatially variable groundwater quality within the basin, with reported concentrations of certain heavy metals approaching or exceeding international drinking water guidelines at specific locations \citep{duncan2018assessment,awuah2016assessment}. Such variability reflects the complex hydrogeological structure of the basin and the diverse potential sources of contamination. The regional evidence from mining- and waste-affected environments in Nigeria further reinforces the need to investigate contaminant distributions within their local hydrogeological contexts rather than assuming that the patterns observed elsewhere apply directly to the Densu Basin \citep{Isah2023,Isah2025a,Isah2025b}.

Notwithstanding these concerns, groundwater quality assessments commonly employ deterministic evaluation frameworks. These include comparisons of measured concentrations against guideline values and the calculation of non-carcinogenic health-risk metrics, such as the Hazard Index (HI), using fixed exposure parameters \citep{donkor2005heavy,means1989risk}. Such approaches provide useful screening-level information and remain valuable for identifying contaminants or locations that warrant further investigation. However, a single calculated HI does not directly quantify the probability that the underlying risk exceeds a specified threshold when concentrations, exposure conditions, and model parameters are uncertain. Moreover, assessing each element separately may not adequately communicate the implications of simultaneous exposure to multiple contaminants. Consequently, deterministic estimates alone may provide an incomplete representation of uncertainty in cumulative health-risk characterisation.

These limitations become particularly consequential when groundwater risk assessments inform public health interventions and water-resource management decisions. Environmental contamination studies increasingly recognise the importance of identifying spatial variability, characterising potential sources, and distinguishing localised hotspots from broader contamination patterns \citep{Chen2018,Isah2025a,Isah2025b}. Nevertheless, spatial identification of contamination does not necessarily establish the probability of a health-risk threshold being exceeded. Concentration-based hotspot analysis and health-risk assessment address related but distinct questions: the former characterises environmental contamination, whereas the latter evaluates the implications of exposure under specified assumptions. Integrating contaminant measurements with an explicit probabilistic health-risk framework can therefore provide complementary information for prioritising monitoring and mitigation.

Recent methodological research in environmental health sciences has called for probabilistic approaches that explicitly integrate uncertainty into risk estimates and provide distributions of potential outcomes rather than single-value indicators \citep{spiegelman2010approaches}. Bayesian statistical methods offer a principled framework for this purpose by representing unknown quantities as probability distributions and updating prior information with observed data to obtain posterior inferences \citep{edition2013bayesian,carlin2008bayesian}. In water-quality and exposure assessment, Bayesian approaches can support uncertainty propagation, estimation of posterior distributions, and probabilistic interpretation of health-risk metrics \citep{yan2019bayesian}. In particular, posterior predictive simulation enables the estimation of the probability that a health-risk measure exceeds a defined threshold, thereby providing information that is not available from a single deterministic estimate.

Despite the growing application of spatial contamination assessment and probabilistic risk analysis, important challenges remain when these approaches are applied to groundwater-dependent communities with limited sampling data. Geospatial characterisation can identify areas with elevated contaminant concentrations, but the resulting maps do not automatically represent uncertainty in cumulative health risks. Conversely, a probabilistic health-risk model can quantify uncertainty in a risk metric but may provide limited spatial decision support unless its outputs are explicitly linked to geographical locations. These distinctions motivate approaches that combine contaminant-specific measurements, probabilistic risk estimation, and spatial representation while clearly communicating the assumptions and limitations associated with each component.

Addressing this gap, the present study adopts a Bayesian probabilistic modelling framework to assess non-carcinogenic health risks associated with exposure to six potentially toxic elements (Fe, Mn, Ni, Pb, Cd, and As) among children in the Densu Basin, Ghana. By integrating field measurements of contaminant concentrations with Bayesian regression and posterior predictive simulation, the analysis propagates uncertainty in model parameters and observations and estimates location-specific probabilities of exceeding the conventional health-index threshold. The resulting probabilities are subsequently represented spatially to support the identification of areas that may warrant further investigation and monitoring. Rather than replacing conventional concentration-based assessment, the proposed framework complements it by providing an uncertainty-aware interpretation of cumulative health-risk estimates.

The contribution of this study lies in the integration of Bayesian probabilistic health-risk estimation with spatial representation of site-level risk probabilities in a groundwater-dependent setting. The analysis focuses on the probability of exceeding a specified non-carcinogenic health-risk threshold rather than relying exclusively on deterministic HI values. This distinction is relevant for decision-making because locations with similar point estimates may differ in the uncertainty associated with their risk characterisation. The resulting framework provides a basis for prioritising groundwater monitoring and further investigation, while recognising that the inferred probabilities are conditional on the observed data, exposure assumptions, statistical model, and spatial interpolation procedure. The study therefore contributes a locally grounded approach to uncertainty-aware groundwater health-risk assessment in data-limited environments.

The remainder of this paper is organised as follows. Section~\ref{sec:study_area} describes the study area and the groundwater quality dataset used in the analysis, including the hydrogeological setting of the Densu River Basin. Section~\ref{sec:hi_framework} outlines the conventional deterministic Health Index framework employed for baseline risk assessment. The Bayesian probabilistic health-risk model is introduced in Section~\ref{sec:bayesian_model}, followed by the posterior predictive risk-assessment methodology in Section~\ref{sec:posterior_predictive}, which quantifies uncertainty in cumulative exposure outcomes. Section~\ref{sec:spatial_risk} presents the spatial risk-mapping approach used to represent site-level probabilistic risk estimates across the basin. The results and their hydrogeological and public-health implications are discussed in Section~\ref{sec:results}. Finally, Section~\ref{sec:conclusion} summarises the key findings, limitations, and policy-relevant implications of the study.

\section{Study Area and Dataset}
\label{sec:study_area}

\subsection{Study Area: Densu Basin}

\begin{figure*}
\includegraphics[width=\textwidth]{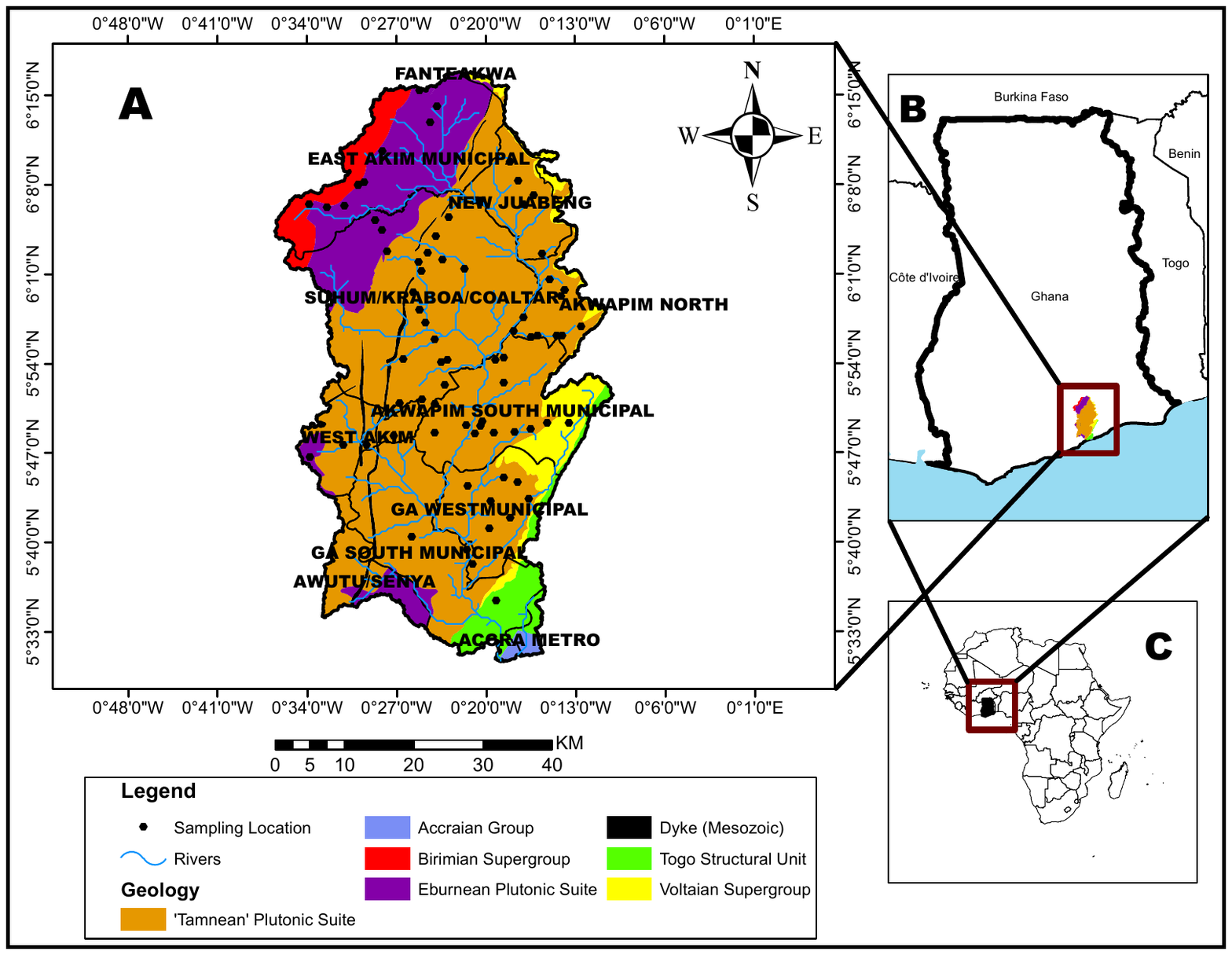}
\caption{Location and geological setting of the Densu River Basin, Ghana. 
(A) Geological map of the Densu Basin showing major lithological units, river networks, administrative boundaries, and groundwater sampling locations used in this study. 
(B) Map of Ghana indicating the position of the Densu Basin within the national context. 
(C) Regional map of Africa highlighting the location of Ghana. 
The spatial distribution of lithological units reflects the geological heterogeneity of the basin, which exerts a strong control on groundwater occurrence, flow pathways, and geochemical characteristics.}
\label{fig:studyAreamap}
\end{figure*}

The Densu River Basin presented in Fig.~\ref{fig:studyAreamap} is located in southern Ghana between latitudes $5^\circ 30'\,\mathrm{N}$ and $6^\circ 20'\,\mathrm{N}$ and longitudes $0^\circ 10'\,\mathrm{W}$ and $0^\circ 35'\,\mathrm{W}$.
The basin drains southwards into the Gulf of Guinea and occupies a strategic hydrological position, supplying surface and groundwater resources to both rural communities and major urban centres, including the Accra Metropolitan Area. Administratively, the basin spans three regions: the Eastern Region ( $\approx 72\%$), the Greater Accra Region ($23\%$), and the Central Region ($5\%$) \citep{nyamekye2016using}. This spatial extent encompasses a diverse range of land-use activities and hydrogeological conditions, making the basin particularly relevant for groundwater quality and health risk investigations.

Hydroclimatic conditions within the Densu Basin are governed by a semi-equatorial climate, resulting in a bimodal rainfall regime with dominant precipitation periods between April--July and September--November.
Mean annual rainfall decreases from the upper catchment towards the southeastern lowlands, reflecting the transition from wetter forested zones to comparatively drier coastal conditions. These seasonal rainfall patterns exert a strong control on groundwater recharge processes, surface groundwater interactions, and the mobilisation and transport of dissolved constituents within the basin’s aquifer systems \citep{amoako2011physico}. Periods of intense rainfall may enhance vertical percolation and contaminant leaching, while prolonged dry seasons can concentrate dissolved species through reduced dilution and increased residence times.

Hydrogeologically, the basin is underlain by a complex assemblage of crystalline basement and sedimentary formations that exert a first-order control on groundwater occurrence, storage, and chemistry. The dominant lithological units include the Birimian Supergroup and associated Tarkwaian sequences, granitoid intrusions of the Eburnean and Tamnean Plutonic Suites, the Accraian Group, the Togo Structural Unit, and portions of the Voltaian Supergroup, together with Mesozoic intrusive dykes \citep{amoako2011physico, akurugu2022groundwater}. Groundwater within these formations is typically hosted in weathered regolith layers and secondary fracture systems, resulting in aquifers of variable transmissivity, limited storage, and pronounced spatial heterogeneity.

The geological complexity of the Densu Basin has important implications for groundwater geochemistry. Basement-derived aquifers are known to exhibit strong lithological controls on major ion composition and trace metal occurrence, with weathering of sulphide-bearing minerals, feldspars, and mafic components contributing to the release of elements such as iron, manganese, and trace metals under favourable geochemical conditions. While not uniformly elevated across the basin, localised zones of enhanced metal mobilisation may arise from the interaction of mineralogical composition, groundwater residence time, redox conditions, and recharge dynamics \citep{akurugu2022groundwater}. These natural controls are further modulated by land-use practices that influence infiltration pathways and subsurface geochemical environments.

Land use within the basin is dominated by agriculture, which includes subsistence food crops such as maize, cassava, plantain, yam, banana, and cocoyam, alongside commercial cultivation of pineapple, oil palm, cocoa, papaya, mango, and citrus \citep{amoako2011physico}. In addition to agriculture, expanding residential development, sand winning, small-scale animal rearing, and salt mining are prevalent in parts of the basin. These activities have contributed to substantial land-cover change and vegetation loss, with associated impacts on runoff generation, soil erosion, sediment transport, and recharge processes. Such alterations can indirectly affect groundwater quality by modifying contaminant loading and transport mechanisms, particularly during high-intensity rainfall events.

Anthropogenic pressures further compound natural hydrogeochemical controls on groundwater quality. The application of agrochemicals, unmanaged waste disposal, urban effluent discharge, and small-scale artisanal mining activities have been identified as potential sources of groundwater contamination in parts of the basin \citep{tay2008groundwater, amoako2011physico}. While the relative contribution of these sources varies spatially, their presence underscores the vulnerability of shallow and fractured aquifers to surface-derived contaminants. In groundwater-dependent communities, particularly those relying on untreated boreholes and hand-dug wells, such vulnerability has direct implications for chronic exposure through drinking water.

Population dependence on groundwater within the Densu Basin remains high, especially in peri-urban and rural settlements where piped surface water supply is intermittent or absent. For children in these communities, groundwater represents a primary exposure pathway for dissolved contaminants, given higher water intake per unit body weight and limited access to alternative water sources. The combination of hydrogeological heterogeneity, spatially variable land-use pressures, and reliance on groundwater for domestic consumption makes the Densu Basin a critical setting for examining uncertainty-aware health risks associated with heavy metal exposure. Accordingly, the basin provides a scientifically and societally relevant context for the probabilistic assessment of groundwater-related health risks among vulnerable populations.

\subsection{Groundwater Dataset and Sampling Design}
\label{sec:data}

\subsubsection{Hydrochemical data collection and laboratory analysis}
\label{sec:collection}

This study is based on a hydrochemical dataset comprising groundwater concentrations of six trace and major metals of public health relevance: lead (Pb), nickel (Ni), cadmium (Cd), iron (Fe), manganese (Mn), and arsenic (As). These elements were selected due to their documented occurrence in crystalline basement aquifers and their recognised implications for drinking water quality and human health within groundwater-dependent settings such as the Densu River Basin.

Groundwater sampling was conducted during a targeted field campaign in January 2020, corresponding to the dry season, when groundwater abstraction for domestic use is typically highest and dilution from surface runoff is limited. A total of 94 sampling locations were selected using a stratified random sampling strategy designed to capture the spatial variability of the basin’s principal aquifer systems and dominant land-use settings. Sampled sources included both mechanised boreholes and hand-dug wells, which together constitute the primary drinking water infrastructure for much of the basin’s population.

All samples were collected in accordance with internationally recognised protocols for trace metal analysis, consistent with the Standard Methods for the Examination of Water and Wastewater. 
Groundwater was collected in pre-cleaned 500~mL low-density polyethene bottles that were rinsed with sample water prior to collection. Immediately after sampling, each sample was preserved by acidification with 1~mL of ultrapure nitric acid (HNO\textsubscript{3}, trace metal grade) to achieve a pH below 2, thereby minimising adsorption of dissolved metals and inhibiting precipitation. Samples were stored in insulated containers at temperatures below \SI{4}{\degreeCelsius} and transported under cold-chain conditions to the central analytical laboratory at the National Nuclear Research Institute, Ghana Atomic Energy Commission.

Prior to instrumental analysis, samples underwent acid digestion to ensure quantification of total recoverable metal concentrations. A 100~mL aliquot of each preserved sample was digested with 5~mL of concentrated HNO\textsubscript{3} on a hot plate at approximately \SI{95}{\degreeCelsius} until the volume was reduced to about 50~mL. The digested samples were then cooled, filtered to remove particulate matter, and diluted to the original volume with deionised water.

Concentrations of As, Cd, Pb, Ni, Fe, and Mn were determined using flame atomic absorption spectrophotometry (FAAS) following established analytical procedures. Quality assurance and quality control measures included the analysis of procedural blanks, duplicate samples, and certified reference materials (CRMs). Recoveries for CRMs ranged between 90\% and 105\%, indicating acceptable analytical performance.

\subsubsection{Univariate statistical characteristics of heavy metals}
\label{sec:univariate}

\begin{figure*}
\includegraphics[width=\textwidth]{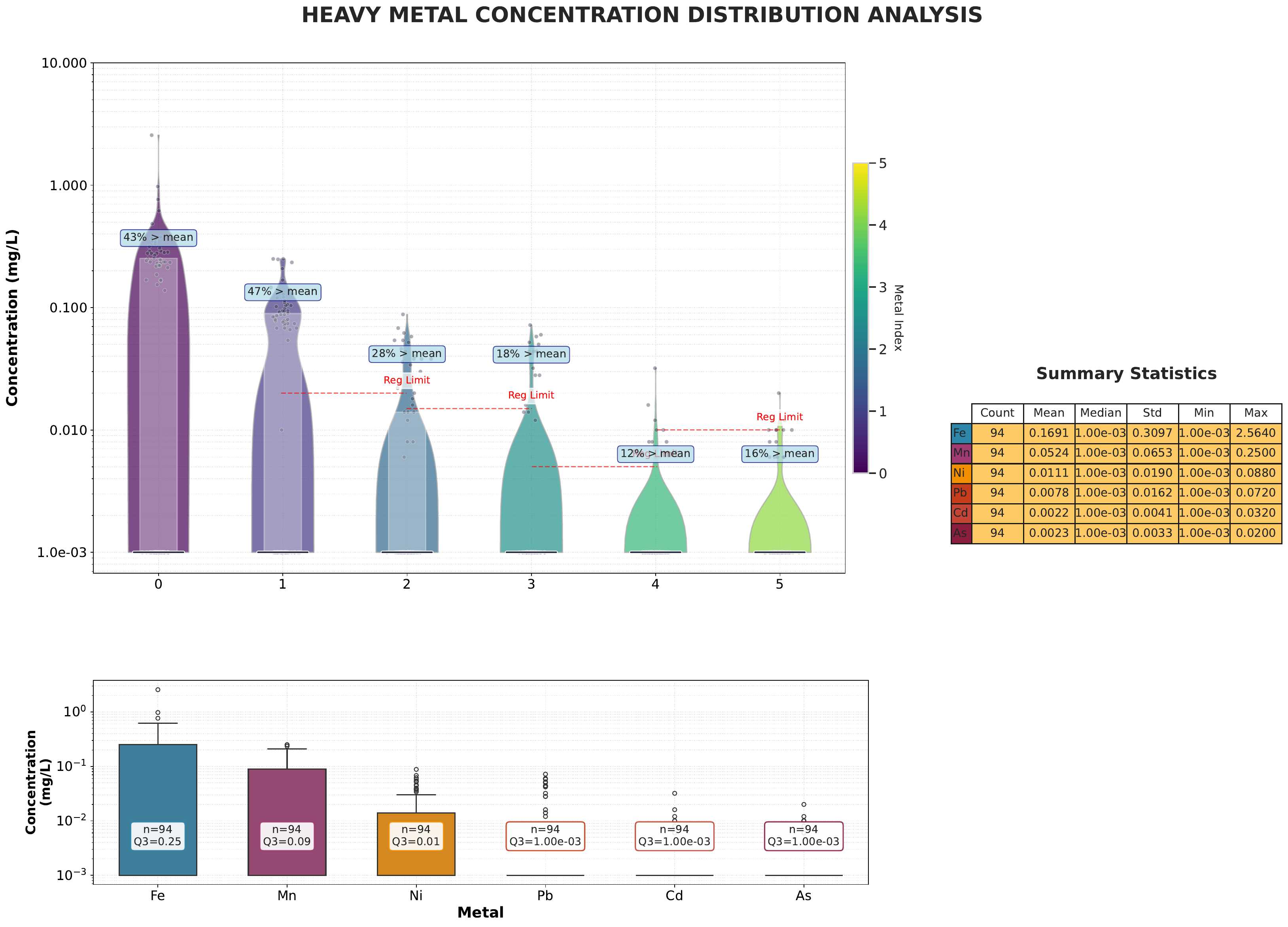}
\caption{Univariate distribution of heavy metal concentrations in groundwater samples from the Densu River Basin. 
Violin plots illustrate the kernel density of concentration values for each metal on a logarithmic scale, with individual observations overlaid as points. 
Box plots displayed below each violin indicate the median, interquartile range, and extreme values. 
Red dashed horizontal lines denote relevant regulatory guideline limits for drinking water. 
The distributions reveal pronounced right skewness, strong concentration at the analytical detection limit, and elevated variability driven by a limited number of high-concentration observations.}
\label{fig:univariate}
\end{figure*}

Univariate characterisation was performed to examine the empirical distribution, dispersion, and concentration structure of heavy metals in groundwater across the Densu River Basin, with explicit attention to how these statistical properties reflect underlying hydrogeological and geochemical controls. Figure~\ref{fig:univariate} integrates three complementary elements: violin plots illustrating distributional density, box plots summarising central tendency and spread, and an embedded summary statistics table quantifying dispersion metrics and regulatory exceedances. 
These components provide a descriptive basis for interpreting groundwater metal behaviour prior to formal exposure or risk modelling.

Groundwater concentrations of Fe, Mn, Ni, Pb, Cd, and As within the Densu Basin display strongly non-symmetric distributions, with a pronounced accumulation of values at low concentrations and elongated upper tails.
This behaviour is clearly depicted in the violin plots, where the widest density occurs near the analytical detection limit (0.001~mg~L$^{-1}$), while progressively narrowing density extends towards higher concentrations. The corresponding box plots reinforce this structure, showing medians fixed at the detection limit and compressed interquartile ranges for most metals. Such distributional patterns are characteristic of crystalline basement aquifers, where groundwater chemistry is governed by spatially heterogeneous weathering profiles, fracture-controlled flow, and localised geochemical micro-environments rather than uniform basin-scale processes \citep{Apambire1997, Edmunds1996}.

Fe displays the broadest concentration range and the most pronounced dispersion, with values spanning more than three orders of magnitude and a coefficient of variation exceeding $180\%$.
The violin plot reveals a dominant low-concentration mode accompanied by a substantial high-concentration tail, while the box plot indicates that variability is driven almost entirely by upper-end values rather than central spread. 
Such extreme variability in the basin is consistent with redox-controlled mobilisation in weathered and fractured basement aquifers, where Fe is released under reducing conditions associated with organic-rich regolith zones and stagnant flow paths \citep{McArthur2012, Smedley2002}.
The annotation indicating that approximately 43\% of Fe observations exceed the mean further underscores the influence of a small number of elevated values in inflating average concentrations, a hallmark of strongly heterogeneous geochemical systems.

Mn exhibits a similar but less extreme distributional structure, with a coefficient of variation exceeding $120\%$.
The co-occurrence of broad Fe and Mn variability is hydrogeochemically meaningful, reflecting their shared redox sensitivity and common association with the reductive dissolution of Fe-Mn oxyhydroxides. The violin density for Mn shows a narrower upper tail than Fe, suggesting more constrained mobilisation, which is consistent with the relative solubility and redox thresholds of Mn-bearing phases in basement-derived aquifers \citep{ zahid2008groundwater, Postma2007, goldberg2006geochemistry, appelo2004geochemistry}.
The observation that nearly half of Mn measurements exceed the mean again indicates dominance of localised enrichment over uniform background conditions.

Ni displays a distinct statistical signature characterised by low central concentrations but pronounced relative variability (CV $\approx$ 170\%). The violin plot indicates a narrow core distribution near the detection limit, punctuated by a discrete upper tail extending beyond regulatory guideline values. The box plot confirms that exceedances are driven by a limited number of samples rather than widespread elevation. In basement terrains such as the Densu Basin, this pattern is consistent with lithological controls, particularly the weathering of mafic minerals or sulphide-bearing assemblages within specific geological units, rather than diffuse anthropogenic loading \citep{alloway2012heavy, kabata2000trace}.
The annotation showing that only a minority of samples exceed the mean highlights the episodic nature of Ni enrichment.

Pb, Cd, and As exhibit the most strongly censored distributions, with medians and interquartile ranges coincident with the detection limit and exceptionally high coefficients of variation exceeding 140--200\%. The violin plots for these elements show minimal density at intermediate concentrations and sparse but distinct upper tails, while the box plots collapse at the lower bound. This statistical structure indicates that background concentrations are generally negligible across most of the basin, with elevated values confined to a small number of locations. In fractured basement settings, such behaviour is typically associated with highly localised sources, including mineralised zones, discrete fracture intersections, or point-scale surface inputs, rather than basin-wide geochemical processes \citep{Lapworth2012, Smedley2002}. The annotated exceedance proportions relative to regulatory limits further emphasise that these metals occur as sporadic anomalies rather than pervasive contaminants.

\subsubsection{Bivariate association analysis}
\label{sec:corr}

To characterise pairwise associations among groundwater metal concentrations, a bivariate analysis was conducted using Spearman’s rank correlation coefficient. This non-parametric measure was selected in preference to moment-based alternatives because the concentration data exhibit pronounced right skewness, extensive left-censoring at the analytical detection limit, and strong influence from extreme values, as demonstrated in the univariate analysis. Under such conditions, rank-based measures provide a more robust representation of monotonic dependence structures than linear correlation coefficients that assume approximate normality and homoscedasticity \citep{smeeton2025applied,helsel2011statistics, conover1999practical}.

Let $\{(X_i, Y_i)\}_{i=1}^n$ denote paired observations of two metal concentrations measured at $n$ sampling locations. Spearman’s correlation coefficient, denoted $\rho_s$ in Eq.~\eqref{eq:spearman_def}, is defined as the Pearson correlation between the rank-transformed variables,
\begin{equation}\label{eq:spearman_def}
\rho_s = \mathrm{corr}\!\left( \mathrm{rank}(X), \mathrm{rank}(Y) \right),
\end{equation}
where tied observations are assigned average ranks.
Statistical significance of the observed correlations was evaluated using permutation-based inference, whereby the empirical null distribution of $\rho_s$ was generated through repeated random reordering of one variable while preserving the marginal distributions. 
Statistical significance of the rank-based associations was evaluated using permutation-based inference, which constructs an empirical null distribution without relying on asymptotic normality assumptions. This approach is particularly well suited to datasets characterised by tied ranks, left-censoring at analytical detection limits, and strongly non-Gaussian distributions \citep{unseld2025permutation, annis2005permutation, good2005permutation, Davison1997}. 

Formally, let $\rho_s^{\text{obs}}$ denote the observed Spearman rank correlation computed from paired observations $\{(X_i, Y_i)\}_{i=1}^n$. Under the null hypothesis of no monotonic association, the joint rank structure is invariant to permutations of one variable. An empirical null distribution is therefore generated via Eq.~\eqref{eq:perm_rho}, by repeatedly permuting the ranks of one variable while holding the other fixed,
\begin{equation}
\label{eq:perm_rho}
\rho_s^{(b)} = \rho_s\!\left(X, \pi_b(Y)\right), \qquad b = 1, \ldots, B,
\end{equation}
where $\pi_b(\cdot)$ denotes the $b$-th random permutation and $B$ is the total number of permutations. The two-sided permutation-based $p$-value is then defined in Eq.~\eqref{eq:perm_pvalue} as
\begin{equation}
\label{eq:perm_pvalue}
p_{\mathrm{perm}} =
\frac{1 + \sum_{b=1}^{B} \mathbb{I}\!\left(\lvert \rho_s^{(b)}\rvert \ge \lvert \rho_s^{\mathrm{obs}}\rvert \right)}{1 + B},
\end{equation}
where $\mathbb{I}(\cdot)$ is the indicator function. Correlations with permutation-based $p$-values below the 5\% significance level were considered statistically significant and are indicated by an asterisk in the correlation matrix.

The resulting correlation structure reveals coherent patterns that are hydrogeochemically meaningful within the context of the Densu River Basin. 
The strongest association is observed between iron (Fe) and manganese (Mn) ($\rho_s = 0.90$, $p < 0.05$), indicating near-synchronous variability across the sampled aquifers. 
Such a high degree of correlation is characteristic of groundwater systems developed within weathered and fractured crystalline basement terrains, where Fe and Mn mobility is primarily controlled by redox conditions rather than by independent source inputs.
In basement aquifers, Fe and Mn commonly occur in the solid phase as ferric and manganic oxyhydroxides under oxic conditions. 
As groundwater residence times increase and dissolved oxygen is progressively depleted along flow paths, these mineral phases become thermodynamically unstable and undergo reductive dissolution. 
This process releases soluble ferrous and manganous species into groundwater, leading to concurrent increases in Fe and Mn concentrations. 
The mobilisation pathways may be schematically represented by the following redox reactions:
\begin{align*}
\mathrm{Fe(III)\text{-}oxyhydroxide} + e^- &\rightarrow \mathrm{Fe^{2+}}, \\
\mathrm{Mn(IV)\text{-}oxide} + 2e^- &\rightarrow \mathrm{Mn^{2+}}.
\end{align*}

\noindent
The close correspondence between Fe and Mn concentrations observed in the correlation matrix therefore reflects a shared geochemical control associated with reducing micro-environments within the aquifer system, such as organic-rich regolith layers, low-permeability fracture zones, or hydraulically stagnant portions of the flow system. 
In the Densu Basin, where groundwater circulation occurs predominantly within weathered profiles and secondary fracture networks, such redox-driven mobilisation is expected to be spatially heterogeneous but process-consistent across sites \citep{McArthur2012, Postma2007}.
The magnitude of the Fe-Mn correlation further suggests that their observed variability is governed predominantly by common hydrogeochemical processes rather than by independent anthropogenic sources. 
This interpretation is consistent with the univariate distribution patterns, which show strong right skewness and extreme coefficients of variation driven by localised enrichment rather than uniform basin-wide elevation. 
Together, these statistical and geochemical indicators support a conceptual model in which Fe and Mn concentrations in the Densu Basin are controlled by redox-sensitive mineral dissolution within the crystalline basement aquifer framework.

\begin{figure*}
\includegraphics[width=\textwidth]{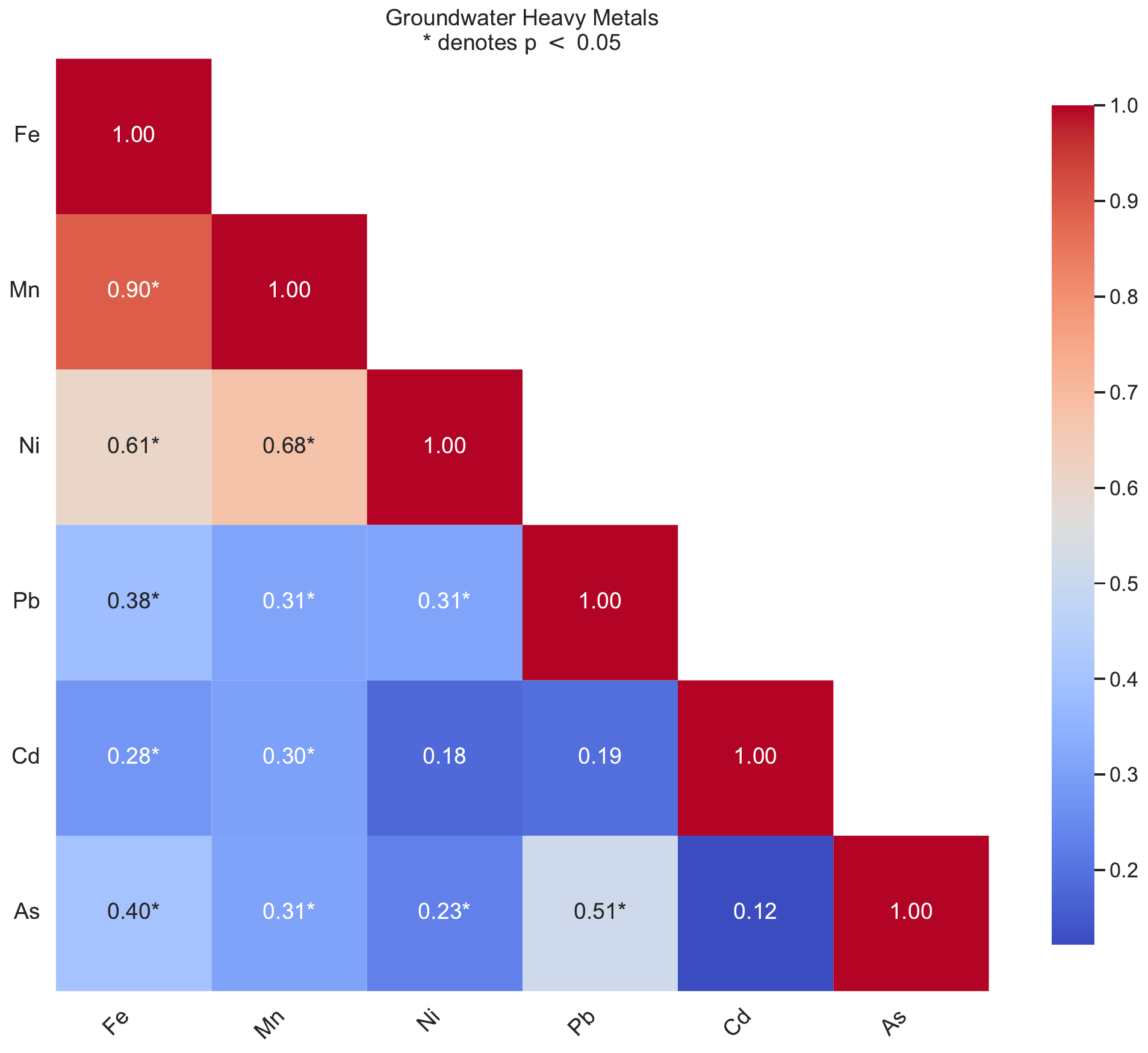}
\caption{Spearman rank correlation matrix of heavy metal concentrations in groundwater samples from the Densu River Basin. 
Correlation coefficients ($\rho_s$) quantify monotonic associations between metal pairs based on ranked concentrations, reducing sensitivity to skewness, outliers, and left-censored observations. 
Statistically significant correlations ($p < 0.05$) are indicated by an asterisk. 
The matrix highlights coherent co-variation among selected metals, reflecting shared hydrogeochemical controls such as lithological composition and redox-sensitive mobilisation within the basin’s crystalline basement aquifers.}
\label{fig:corrmatrix}
\end{figure*}

Ni exhibits moderate to strong positive correlations with both Fe ($\rho_s = 0.61$) and Mn ($\rho_s = 0.68$), implying partial co-mobilisation. In basement terrains, Ni is commonly associated with mafic and ultramafic minerals and may be released during the weathering of ferromagnesian phases or adsorbed onto Fe–Mn oxyhydroxides that subsequently dissolve under reducing conditions. The observed correlations therefore suggest that Ni enrichment in the basin is linked, at least in part, to the same lithological and geochemical processes influencing Fe and Mn, rather than arising solely from diffuse surface inputs.

Pb and As display a moderate positive association ($\rho_s = 0.51$, $p < 0.05$), despite their generally low median concentrations. This relationship is consistent with their known affinity for adsorption onto Fe-bearing mineral surfaces and subsequent release during changes in redox conditions or pH.
In fractured basement aquifers, such co-variation often reflects mobilisation along discrete flow paths intersecting mineralised zones or iron-rich weathering horizons, rather than basin-wide geochemical uniformity.

Cd shows weak and largely non-significant correlations with the other metals, indicating a decoupled spatial distribution. This statistical independence suggests that Cd concentrations may be influenced by more heterogeneous or site-specific factors, including localised anthropogenic inputs or distinct mineralogical sources, rather than the dominant redox-driven processes controlling Fe and Mn. Similar patterns have been reported in basement aquifers where Cd occurrence is sporadic and poorly correlated with major redox-sensitive elements.

\section{Conventional Health Risk Assessment Framework}
\label{sec:hi_framework}

\subsection{Conceptual Basis of the Health Index}
\label{sec:hi_concept}

In groundwater quality assessment, the Health Index (HI) is widely employed as a screening-level metric for evaluating potential non-carcinogenic risks associated with chronic exposure to multiple contaminants through drinking water. The index is designed to condense information from several exposure pathways into a single dimensionless quantity, thereby facilitating comparison across locations and against established benchmarks for acceptable risk \citep{maurya2025hydrochemical, okyere2025human, assessment2009science}. Within this framework, the HI does not seek to predict clinical outcomes, but rather to provide an integrative indicator of relative concern that can inform prioritisation and further investigation.

The conceptual foundation of the HI rests on the principle of dose additivity, whereby the combined effect of multiple contaminants is approximated as the sum of their individual hazard contributions. This approach is rooted in regulatory risk assessment practice and has been adopted extensively in groundwater studies due to its transparency, computational simplicity, and compatibility with limited datasets \citep{simpson2014assessing}. 
As a scalar diagnostic variable, the HI enables spatial comparison of aggregate exposure conditions and supports preliminary assessment of whether contaminant levels may warrant closer scrutiny.

Children are commonly considered as a distinct exposure group within the HI framework owing to physiological and behavioural characteristics that differentiate their exposure profiles from those of adults. Higher water intake per unit body mass, ongoing developmental processes, and greater sensitivity to toxicological stressors render children particularly relevant for conservative risk screening in groundwater-dependent communities \citep{ferguson2017review, kavcar2009health, williams2003current, weaver1998approaches}. Accordingly, the HI(children) formulation is widely used in environmental health studies as a protective indicator of potential vulnerability.

\subsection{Computation of the Health Index for Children}
\label{sec:hi_computation}

The conventional computation of the Health Index for children, denoted
$\mathrm{HI}_{\text{child}}$, involves aggregating contaminant-specific
hazard quotients (HQs). For each contaminant $j$, the hazard quotient is
defined as the ratio of the estimated average daily dose to a reference
dose considered to be without appreciable risk of adverse effects:
\begin{equation}
\mathrm{HQ}_j = \frac{\mathrm{ADD}_j}{\mathrm{RfD}_j},
\end{equation}
where $\mathrm{ADD}_j$ is the average daily dose
(mg\,kg$^{-1}$\,day$^{-1}$) and $\mathrm{RfD}_j$ is the oral reference
dose for contaminant $j$ \citep{EPA_EFH_2011}.

For ingestion of contaminated groundwater, the average daily dose for
children is conventionally expressed as
\begin{equation}
\mathrm{ADD}_j =
\frac{C_j \times IR \times EF \times ED}{BW \times AT},
\end{equation}
where $C_j$ denotes the measured concentration of contaminant $j$ in
drinking water (mg\,L$^{-1}$), $IR$ is the ingestion rate (L\,day$^{-1}$),
$EF$ is the exposure frequency (days\,year$^{-1}$), $ED$ is the exposure
duration (years), $BW$ is body weight (kg), and $AT$ is the averaging
time (days). Parameter values for children are typically obtained from
standard exposure-factor compilations and reflect age-specific intake
patterns and body mass distributions.

The overall Health Index for children is calculated by summing the hazard
quotients across the $M$ contaminants considered:
\begin{equation}
\mathrm{HI}_{\text{child}} = \sum_{j=1}^{M} \mathrm{HQ}_j.
\end{equation}
Values of $\mathrm{HI}_{\text{child}} < 1$ are conventionally interpreted
as indicating a lower likelihood of appreciable non-carcinogenic concern
under the specified exposure assumptions, whereas values
$\mathrm{HI}_{\text{child}} \geq 1$ indicate potential concern and warrant
further assessment or management consideration. The threshold is a
screening criterion rather than a direct measure of the probability or
severity of an adverse health outcome.

Although the Health Index provides a widely used screening measure, its
interpretation depends on the assumptions underlying the exposure
assessment. In conventional applications, contaminant concentrations and
exposure parameters are often represented by fixed values, which may not
adequately reflect spatial variability, temporal fluctuations, or
measurement uncertainty. In heterogeneous hydrogeological settings, such
as crystalline basement aquifers, these simplifications may obscure
local variation in groundwater quality and potential exposure.

The HI formulation also assumes additive contributions from individual
hazard quotients. This is consistent with screening-level risk assessment
practice, but it does not explicitly represent interactions among
contaminants or dependence arising from shared hydrogeochemical controls,
such as redox-driven mobilisation in groundwater systems
\citep{altenburger2013simplifying}.

An important consideration for the present study is that the Health Index
is itself calculated from the measured contaminant concentrations and
the specified exposure parameters. Consequently, any subsequent
regression of HI against those same concentrations involves mathematical
dependence between the response and its predictors. The regression
should therefore not be interpreted as establishing independent
relationships between contaminant concentrations and health outcomes.
Instead, its results must be understood in relation to the construction
of HI and the assumptions of the statistical model.

The Health Index is intended primarily as a screening indicator of
potential non-carcinogenic concern, rather than a predictive measure of
clinical outcomes. Its deterministic formulation does not, by itself,
quantify uncertainty in the estimated risk or the probability that the
screening threshold will be exceeded. These limitations motivate the
probabilistic assessment developed in this study, in which uncertainty
in the model parameters and residual variability is propagated into
estimates of HI exceedance. The resulting probabilities are conditional
on the specified model and exposure assumptions and should not be
interpreted as probabilities of observed adverse health effects.

\section{Bayesian Health Risk Model}
\label{sec:bayesian_model}

Health risk assessments based on groundwater quality involve uncertainty
arising from variation in contaminant concentrations, measurement error,
exposure assumptions, and the limitations of the risk-assessment model.
This study adopts a Bayesian regression framework to represent uncertainty
in model parameters through posterior probability distributions and
propagate that uncertainty into estimates of Health Index (HI) exceedance
\citep{edition2013bayesian, carlin2008bayesian, gelman1995bayesian}.
The objective is to characterise uncertainty in the calculated screening
index under the specified statistical model, rather than to predict
independently observed clinical health outcomes.

The groundwater metal concentrations used as model inputs reflect the
combined influence of hydrogeological processes operating within the
study area, including aquifer lithology, geochemical conditions, and
groundwater flow pathways. These processes are not modelled explicitly;
inference is instead conditioned on the observed concentrations and the
constructed Health Index values.

Let $\mathrm{HI}_i$ denote the Health Index for children at sampling
location $i$, calculated using the formulation in
Section~\ref{sec:hi_computation}. The second-stage regression represents
this index as a linear function of the standardised metal concentrations
with an additive stochastic component:
\begin{equation}
\mathrm{HI}_i
=
\alpha
+
\sum_{j=1}^{6} \beta_j X_{ij}
+
\varepsilon_i,
\label{eq:bayesian_hi_model}
\end{equation}
where $\alpha$ is the intercept, $\beta_j$ is the regression coefficient
for metal $j$, and $X_{ij}$ denotes the standardised concentration of
metal $j$ at location $i$. The six metals considered are Fe, Mn, Ni, Pb,
Cd, and As. Standardisation is performed by subtracting the sample mean
and dividing by the sample standard deviation for each contaminant.
This transformation places the predictors on comparable scales and
supports numerical stability during estimation.

The residual term represents variation in HI not captured by the
specified linear predictor. Such variation may reflect departures from
the assumed linear relationship, variation in the constructed index
associated with exposure assumptions, and other sources of unexplained
variability. The residuals are assumed to follow a Gaussian distribution:
\begin{equation}
\varepsilon_i \sim \mathcal{N}(0, \sigma^2),
\label{eq:residual_distribution}
\end{equation}
which implies the likelihood
\begin{equation}
p(\mathbf{HI} \mid \alpha, \boldsymbol{\beta}, \sigma, \mathbf{X})
=
\prod_{i=1}^{n}
\mathcal{N}
\left(
\mathrm{HI}_i \mid
\alpha + \sum_{j=1}^{6} \beta_j X_{ij},
\sigma^2
\right),
\label{eq:likelihood}
\end{equation}
where $n$ is the number of sampling locations.

The relationship specified in Eq.~\eqref{eq:bayesian_hi_model} requires
careful interpretation because HI is calculated from the same metal
concentrations used as predictors. The response and predictors are
therefore not independent quantities, and the regression cannot be
regarded as an independent test of contaminant effects or as evidence
of causal relationships with health outcomes. In particular, the
estimated coefficients describe conditional associations within the
specified regression representation of the calculated HI; they do not
necessarily quantify the independent contribution of each metal to
health risk beyond its contribution to the original HI calculation.

The rationale for this second-stage model is to represent the
calculated HI through a probabilistic regression and propagate parameter
uncertainty and residual variability into the subsequent exceedance
assessment. Its additional output is a model-conditional distribution
of HI values and corresponding probabilities of exceeding the screening
threshold, rather than a new or independent measure of health risk.
These quantities depend on the assumed linear predictor, likelihood,
prior distributions, observed concentrations, and exposure assumptions.
The probabilistic regression should therefore be interpreted as an
uncertainty-propagation framework, not as evidence that the HI
calculation itself has been independently validated or that the model
improves on direct probabilistic modelling of HI without a comparative
analysis.

Prior distributions are specified to complete the Bayesian model.
Weakly informative normal priors are assigned to the intercept and
regression coefficients:
\begin{equation}
\alpha \sim \mathcal{N}(0,1), \qquad
\beta_j \sim \mathcal{N}(0,1), \quad j = 1,\ldots,6,
\label{eq:priors_alpha_beta}
\end{equation}
reflecting prior uncertainty centred on zero while permitting
substantial deviations when supported by the data. The residual standard
deviation is assigned a half-normal prior:
\begin{equation}
\sigma \sim \mathcal{HN}(1),
\label{eq:prior_sigma}
\end{equation}
which constrains $\sigma$ to positive values and provides regularisation
without imposing a restrictive assumption on unexplained variability
\citep{Browne2006, gelman2006prior, gustafson2006conservative}.

Combining the likelihood in Eq.~\eqref{eq:likelihood} with the prior
distributions in Eqs.~\eqref{eq:priors_alpha_beta} and
\eqref{eq:prior_sigma} yields the joint posterior distribution:
\begin{equation}
p(\alpha, \boldsymbol{\beta}, \sigma \mid \mathbf{HI}, \mathbf{X})
\propto
p(\mathbf{HI} \mid \alpha, \boldsymbol{\beta}, \sigma, \mathbf{X})
p(\alpha)
\prod_{j=1}^{6} p(\beta_j)
p(\sigma).
\label{eq:posterior}
\end{equation}
This posterior distribution defines the target for inference. Because
the likelihood couples the parameters, the posterior is not available
in a convenient closed form and is estimated numerically.

\begin{figure*}
\centering
\includegraphics[width=\textwidth]{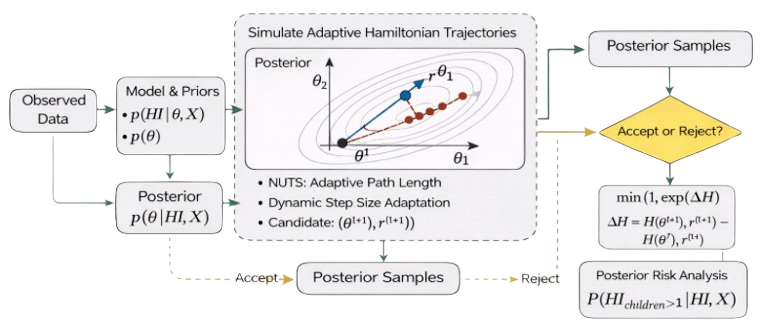}
\caption{Schematic illustration of the Adaptive Hamiltonian Monte Carlo (HMC)
sampling framework employed for Bayesian health risk assessment. Observed
groundwater contaminant data are combined with prior distributions to define the
posterior density of the model parameters
$\boldsymbol{\theta} = (\alpha, \boldsymbol{\beta}, \sigma)$ as specified in
Eq.~\eqref{eq:posterior}. Posterior exploration is performed using the No-U-Turn
Sampler (NUTS), which adaptively simulates Hamiltonian trajectories by
dynamically tuning the step size and trajectory length. Candidate states
$(\boldsymbol{\theta}^{(t+1)}, \mathbf{r}^{(t+1)})$ are accepted with probability
$\min\{1, \exp(\Delta H)\}$, where $\Delta H$ denotes the change in Hamiltonian
energy between successive states. Accepted samples are subsequently propagated
through the posterior predictive model in Eq.~\eqref{eq:posterior_predictive} to
estimate the probability of exceedance of the health risk threshold defined in
Eq.~\eqref{eq:prob_hi_exceedance}.}
\label{fig:hmc}
\end{figure*}

Posterior inference is carried out using Markov Chain Monte Carlo (MCMC)
sampling. Specifically, the No-U-Turn Sampler (NUTS), an adaptive variant
of Hamiltonian Monte Carlo, is employed to explore the posterior
distribution defined by Eq.~\eqref{eq:posterior}
\citep{hase2025probabilistic, marsden2025randomized,
oberdorster2025accelerated, betancourt2017conceptual,
hoffman2014no}.
As illustrated in Fig.~\ref{fig:hmc}, NUTS uses gradient information
from the log-posterior to simulate Hamiltonian dynamics, allowing
proposals to move efficiently through regions of high posterior density.
Adaptation of the step size and trajectory length reduces sensitivity
to tuning parameters and can mitigate the random-walk behaviour
associated with simpler MCMC algorithms. Convergence is assessed using
the potential scale reduction factor ($\hat{R}$), effective sample size,
and visual inspection of trace plots.


\section{Posterior Predictive Risk Assessment}
\label{sec:posterior_predictive}

Posterior samples obtained from Section~\ref{sec:bayesian_model} are propagated into predicted Health Index values to quantify uncertainty in health risk outcomes. For each posterior draw $k$, replicated Health Index values are generated according to

\begin{equation}
\mathrm{HI}_i^{(k)}
\sim
\mathcal{N}
\left(
\alpha^{(k)}
+
\sum_{j=1}^{6} \beta_j^{(k)} X_{ij},
\sigma^{(k)2}
\right),
\label{eq:posterior_predictive}
\end{equation}

\noindent thereby yielding a posterior predictive distribution that reflects uncertainty arising from both parameter estimation and unexplained variability in the Health Index.

The posterior predictive distribution is used here to propagate uncertainty in the model parameters and residual variability into the assessment of Health Index exceedance. Its principal role is to quantify uncertainty associated with the specified risk-assessment framework, rather than to establish independent out-of-sample predictive performance. Accordingly, the resulting exceedance probabilities are interpreted conditionally on the observed groundwater data, the assumed regression structure, and the exposure assumptions underlying the Health Index.

In regulatory health risk assessment, values of the Health Index exceeding unity are commonly interpreted as indicating potential non-carcinogenic concern. Accordingly, the primary risk metric considered in this study is the posterior probability of exceedance,

\begin{equation}
P\!\left(
\mathrm{HI}_{\mathrm{children}} > 1
\mid \mathbf{HI}, \mathbf{X}
\right),
\label{eq:prob_hi_exceedance}
\end{equation}

\noindent which represents the probability, conditional on the observed data and model assumptions, that cumulative exposure levels for children exceed the accepted safety threshold. This quantity provides a probabilistic characterisation of health risk, rather than a binary classification based on a single deterministic estimate.

To further elucidate the contribution of individual contaminants to cumulative risk, the posterior probability that each regression coefficient is positive is computed as

\begin{equation}
P(\beta_j > 0)
=
\frac{1}{N}
\sum_{k=1}^{N}
\mathbb{I}\!\left( \beta_j^{(k)} > 0 \right),
\label{eq:beta_positive_probability}
\end{equation}

\noindent where $\beta_j^{(k)}$ denotes the $k$th posterior draw for metal $j$ and $\mathbb{I}(\cdot)$ is the indicator function. When interpreted alongside the exceedance probability in Eq.~\eqref{eq:prob_hi_exceedance}, this measure provides an uncertainty-aware assessment of the likelihood of exceeding the Health Index threshold and the direction of the association between individual metals and the Health Index. A positive coefficient indicates a positive conditional association within the specified regression model; it does not, by itself, establish a causal contribution to cumulative health risk.

\section{Spatial Risk Mapping}
\label{sec:spatial_risk}

The final stage of the proposed framework translates posterior predictive health
risk estimates into a spatially continuous representation of risk across the
Densu Basin. While the Bayesian model yields probabilistic estimates of Health
Index exceedance at discrete sampling locations, environmental management and
public health decision-making require spatially explicit information that
extends beyond individual points. Spatial risk mapping therefore, provides a
methodological bridge between Bayesian inference and geographically targeted
risk assessment.

\subsection{Site-Level Posterior Risk Estimation}

Posterior predictive samples obtained from Section~\ref{sec:posterior_predictive} were used to quantify the probability of
health risk exceedance at each sampling location. For a given location $i$, the
posterior risk probability was computed as

\begin{equation}
R_i
=
P\!\left(
\mathrm{HI}_i > 1 \mid \mathbf{HI}, \mathbf{X}
\right)
=
\frac{1}{N}
\sum_{k=1}^{N}
\mathbb{I}\!\left(
\mathrm{HI}_i^{(k)} > 1
\right),
\label{eq:site_risk_probability}
\end{equation}

\noindent
where $\mathrm{HI}_i^{(k)}$ denotes the $k$th posterior predictive draw of the
Health Index at location $i$, $N$ is the total number of posterior samples across
all chains, and $\mathbb{I}(\cdot)$ is the indicator function. As defined in
Eq.~\eqref{eq:site_risk_probability}, $R_i$ represents the posterior probability
that cumulative exposure at location $i$ exceeds the non-carcinogenic risk
threshold.
The resulting set $\{R_i\}_{i=1}^{n}$ constitutes discrete, uncertainty-aware
estimates of health risk at observed locations. These values, bounded on the
interval $[0,1]$, form the input for spatial interpolation and enable the
subsequent construction of continuous posterior risk surfaces.

\subsection{Spatial Interpolation of Posterior Risk}
\label{sec:spatial_interpolation}

To represent the spatial distribution of health risk across the Densu
Basin, the site-specific posterior risk probabilities $\{R_i\}$ were
interpolated onto a regular spatial grid. Inverse Distance Weighting
(IDW) was used to estimate risk values at grid locations between the
observed groundwater sampling points. This deterministic interpolation
method estimates the value at an unsampled location as a weighted
average of the observed values, assigning greater weight to sampling
locations that are geographically closer to the prediction point.
Consequently, the interpolated surface reflects the spatial pattern
of the estimated posterior risk probabilities while retaining the
observed site-level estimates as the underlying data.

The choice of IDW defines how information from the sampling locations
is transferred to the intervening grid cells. Unlike geostatistical
methods such as ordinary kriging, IDW does not require estimation of
a variogram or specification of a spatial covariance structure.
However, it also does not explicitly model spatial autocorrelation or
provide a model-based estimate of interpolation uncertainty. Its use
in this study is therefore restricted to constructing a continuous
spatial representation of the site-level posterior risk estimates;
the interpolated values are not treated as independently observed
risk probabilities or as evidence of validated predictive accuracy
at unsampled locations. The spatial surface is interpreted in
conjunction with the original sampling locations and their associated
posterior risk estimates.

Formally, the interpolated posterior risk at grid location $\mathbf{s}$
is given by

\begin{equation}
\hat{R}(\mathbf{s})
=
\frac{
\sum_{i=1}^{n} w_i(\mathbf{s}) \, R_i
}{
\sum_{i=1}^{n} w_i(\mathbf{s})
},
\qquad
w_i(\mathbf{s}) = \frac{1}{d(\mathbf{s}, \mathbf{s}_i)^p},
\label{eq:idw_risk}
\end{equation}

\noindent
where $d(\mathbf{s}, \mathbf{s}_i)$ denotes the Euclidean distance between the
prediction location $\mathbf{s}$ and observation location $\mathbf{s}_i$, and
$p$ is a positive distance decay exponent controlling the rate at which the
influence of observations decreases with distance. As shown in
Eq.~\eqref{eq:idw_risk}, the interpolated risk surface is a convex combination of
site-level posterior risk probabilities.

The selection of IDW is motivated by both methodological and data-related
considerations. First, the spatial density and distribution of groundwater
sampling points within the basin are insufficient to reliably estimate spatial
covariance structures required for higher-order geostatistical approaches such
as kriging \citep{cressie2015statistics}. Second, IDW imposes minimal assumptions
on the underlying spatial process, thereby reducing the risk of artefacts
associated with over-parameterisation under sparse sampling conditions
\citep{li2008review}. Third, the use of IDW ensures methodological consistency
with the interpolation framework applied to individual heavy metal concentrations, facilitating coherent comparison between contaminant
distributions and posterior risk patterns.
To constrain interpolation to the spatial support of the available data, risk
estimation was restricted to locations within the basin boundary and masked
based on proximity to the nearest observation. This procedure limits
extrapolation beyond empirically informed regions and ensures that the resulting
risk surface remains anchored to observed posterior risk estimates.

The computational procedure used to implement IDW interpolation of posterior
risk probabilities is summarised in Algorithm~\ref{alg:idw_risk}, which details
the calculation of distance-based weights and the construction of the
interpolated risk field on the prediction grid.

\begin{algorithm}
\caption{Inverse Distance Weighting (IDW) Interpolation of Posterior Risk}
\label{alg:idw_risk}
\begin{algorithmic}[1]
\Require
Observed locations $\{\mathbf{s}_i\}_{i=1}^{n}$,
posterior risk probabilities $\{R_i\}_{i=1}^{n}$,
prediction grid $\{\mathbf{s}_g\}_{g=1}^{G}$,
distance exponent $p > 0$
\Ensure
Interpolated risk surface $\{\hat{R}(\mathbf{s}_g)\}_{g=1}^{G}$

\For{$g = 1$ to $G$}
    \State Initialise numerator $\gets 0$
    \State Initialise denominator $\gets 0$
    \For{$i = 1$ to $n$}
        \State Compute distance $d \gets \lVert \mathbf{s}_g - \mathbf{s}_i \rVert$
        \If{$d > 0$}
            \State Compute weight $w \gets d^{-p}$
            \State Update numerator $\gets$ numerator $+ w \cdot R_i$
            \State Update denominator $\gets$ denominator $+ w$
        \EndIf
    \EndFor
    \If{denominator $> 0$}
        \State $\hat{R}(\mathbf{s}_g) \gets$ numerator / denominator
    \Else
        \State $\hat{R}(\mathbf{s}_g) \gets \text{NaN}$
    \EndIf
\EndFor
\end{algorithmic}
\end{algorithm}

\section{Results and Discussion}
\label{sec:results}

\subsection{Bayesian Diagnostic Assessment and Parameter Inference}
\label{sec:Diagnostic}

\begin{figure*}
\centering
\includegraphics[width=\textwidth]{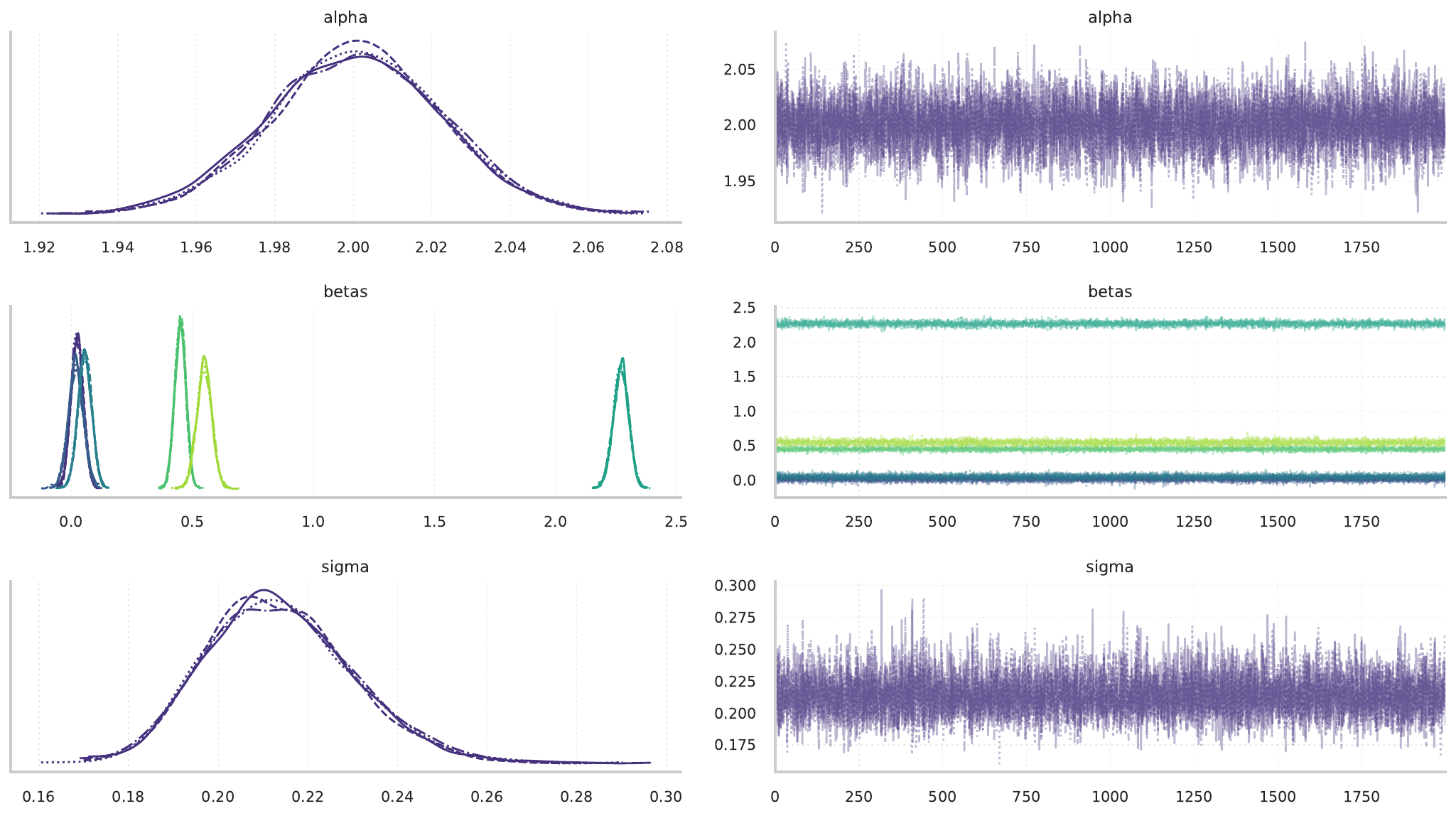}
\caption{Trace plots and marginal posterior distributions for the intercept
($\alpha$), metal-specific regression coefficients ($\beta_i$), and residual
standard deviation ($\sigma$) from the Bayesian health risk model. The stable
trace behaviour and unimodal posterior densities indicate satisfactory
convergence and effective exploration of the posterior parameter space.}
\label{fig:trace_plot}
\end{figure*}

The adequacy of the Bayesian health risk model was first assessed through diagnostic evaluation of the MCMC output. Figure~\ref{fig:trace_plot} shows the trace plots and marginal posterior distributions for the intercept ($\alpha$), the metal-specific regression coefficients ($\beta_i$), and the residual standard deviation ($\sigma$). Across all parameters, the trace plots exhibit stable behaviour, with no discernible trends, abrupt shifts, or evident signs of multimodality. The chains fluctuate around well-defined means, suggesting satisfactory mixing and exploration of the posterior distribution.

These diagnostic patterns provide evidence that the sampling algorithm has explored the posterior distribution without obvious sampling pathologies. In particular, the absence of persistent drift or abrupt changes supports the reliability of the posterior summaries used in the subsequent risk assessment. However, convergence diagnostics assess the behaviour of the sampling algorithm and do not, on their own, establish the adequacy of the assumed likelihood or guarantee accurate predictions at unobserved locations. The diagnostic evidence is therefore interpreted in relation to the principal objective of this study, which is to quantify uncertainty in groundwater-related Health Index exceedance.

The posterior distribution of the intercept $\alpha$ is unimodal and relatively narrow, indicating a comparatively well-constrained estimate of the baseline Health Index under the specified regression model and after standardisation of the predictor variables. In this model, $\alpha$ represents the expected baseline level of the Health Index when the standardised predictor variables are at their reference values. It should not be interpreted as a direct measure of the effects of specific unmeasured exposure pathways, such as dietary intake or household water-handling practices, unless these influences are explicitly represented in the model.

The posterior distribution of the residual standard deviation $\sigma$ characterises the variability in the child Health Index that remains unexplained by the modelled metal concentrations. From a hydrogeological perspective, this residual variability may reflect local lithological heterogeneity, differences in redox conditions, variation in groundwater residence time, and site-specific exposure pathways. These processes may contribute to spatial variation that is not fully captured by the measured predictors. The posterior distribution of $\sigma$ therefore provides information about residual variability conditional on the specified model, rather than, by itself, establishing whether the model is adequate or inadequate. Such variability is consistent with the complexity of groundwater systems, in which microscale geochemical processes and local exposure conditions may not be fully resolved using regional-scale data \citep{edmunds2008natural, jordana2004natural}.

\begin{figure*}
\centering
\includegraphics[width=\textwidth]{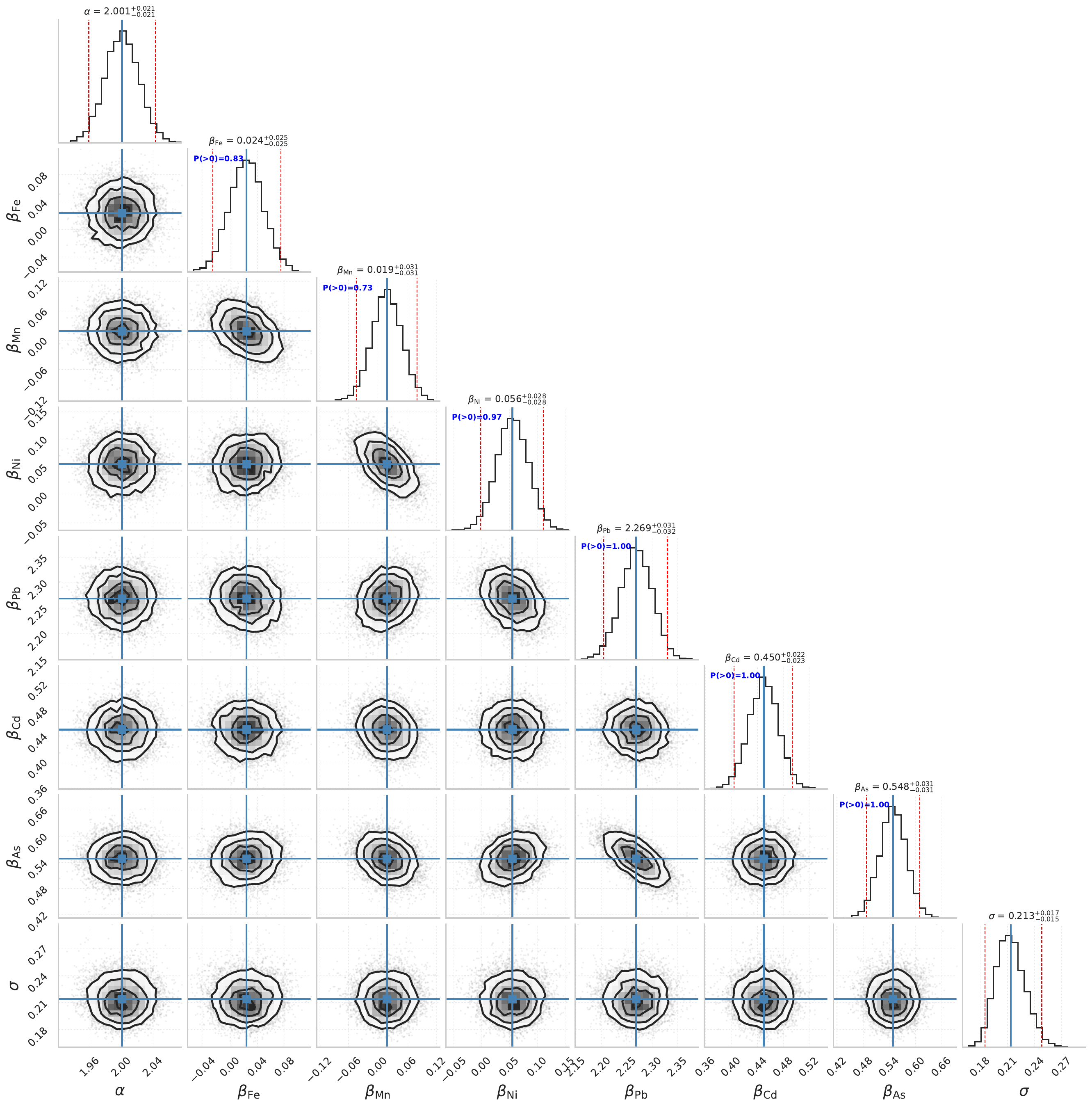}
\caption{Corner plot showing the joint posterior distribution of model parameters,
including the intercept ($\alpha$), regression coefficients for individual heavy
metals ($\beta_{\mathrm{Fe}}$, $\beta_{\mathrm{Mn}}$, $\beta_{\mathrm{Ni}}$,
$\beta_{\mathrm{Pb}}$, $\beta_{\mathrm{Cd}}$, $\beta_{\mathrm{As}}$), and the
residual error term ($\sigma$). Diagonal panels display marginal posterior
distributions with 95\% credible intervals (red dashed lines), while off-diagonal
panels illustrate pairwise parameter dependencies. Annotated values denote the
posterior probability of a positive effect, $P(\beta_i > 0)$.}
\label{fig:corner_plot_all_parameters_annotated}
\end{figure*}

The joint posterior structure of the model parameters is illustrated in
Fig.~\ref{fig:corner_plot_all_parameters_annotated}. Importantly, this figure
depicts posterior distributions of the regression parameters themselves rather
than the observed metal concentrations. The diagonal panels show marginal
posterior densities, while the off-diagonal panels reveal pairwise dependencies
between parameters. The predominantly elliptical contours observed for most
parameter pairs indicate weak to moderate posterior correlations, suggesting
that the model is able to disentangle the individual contributions of co-occurring
metals. This is particularly relevant in hydrogeological settings where multiple
trace elements may be mobilised simultaneously through shared processes such as
mineral weathering, reductive dissolution, and groundwater–rock interaction along
common flow paths.

Several regression coefficients exhibit posterior distributions that are clearly
shifted away from zero, whereas others remain centred closer to zero with wider
credible intervals. To translate these parameter estimates into an interpretable
measure of health relevance, the posterior probability of a positive effect,
$P(\beta_i > 0)$, was computed for each metal. This quantity represents the
Bayesian degree of belief that increasing concentrations of a given metal are
associated with an increase in the child Health Index, conditional on the
observed data and model assumptions. By reporting $P(\beta_i > 0)$ alongside
95\% credible intervals, the analysis explicitly links statistical inference with
environmental interpretation in a transparent and reproducible manner.

Based on posterior effect magnitude and $P(\beta_i > 0)$, the metals can be
ranked according to their inferred contribution to cumulative health risk as
Pb $>$ As $>$ Cd $>$ Ni $>$ Mn $>$ Fe. This ranking reflects relative strength of
association within the probabilistic framework rather than absolute toxicity.
The dominant role of Pb is consistent with its well-documented neurotoxic effects
in children and its persistence in groundwater systems influenced by
anthropogenic activities, including informal mining, waste disposal, and ageing
distribution infrastructure \citep{Fewtrell2005}. From a hydrogeological
perspective, lead mobilisation is often enhanced under mildly acidic conditions
and in aquifers affected by urban and peri-urban land use, which characterise
parts of the study area.

Arsenic also exhibits a high posterior probability of positive association with the Health Index, consistent with its predominantly geogenic origin in many groundwater systems. Reductive dissolution of iron oxyhydroxides is a well-established mechanism for arsenic release, particularly under low-oxygen conditions common in deeper or poorly flushed aquifer zones. The joint posterior
structure involving $\beta_{\mathrm{As}}$ and $\beta_{\mathrm{Fe}}$ provides
additional support for this hydrogeochemical linkage, even though iron itself
shows a comparatively weaker direct association with health risk.

Cadmium and Ni display moderate posterior probabilities, suggesting
context-dependent contributions that likely reflect mixed geogenic and
anthropogenic sources, including mineral weathering, agricultural inputs, and
localised industrial activities. Their wider credible intervals indicate greater
spatial variability in source intensity and transport processes, consistent with
heterogeneous aquifer conditions and variable land-use pressures. In contrast,
Mn and Fe, while frequently elevated in groundwater from a geochemical
standpoint, exhibit lower posterior probabilities of positive association with
the Health Index. This distinction underscores the importance of separating
geochemical abundance from toxicological relevance, particularly in probabilistic
frameworks that explicitly account for uncertainty and co-exposure.

\subsection{Spatial Distribution of Metal Concentrations and Posterior Health Risk}
\label{sec:spatial_results}

\begin{figure*}
\centering
\includegraphics[width=\textwidth]{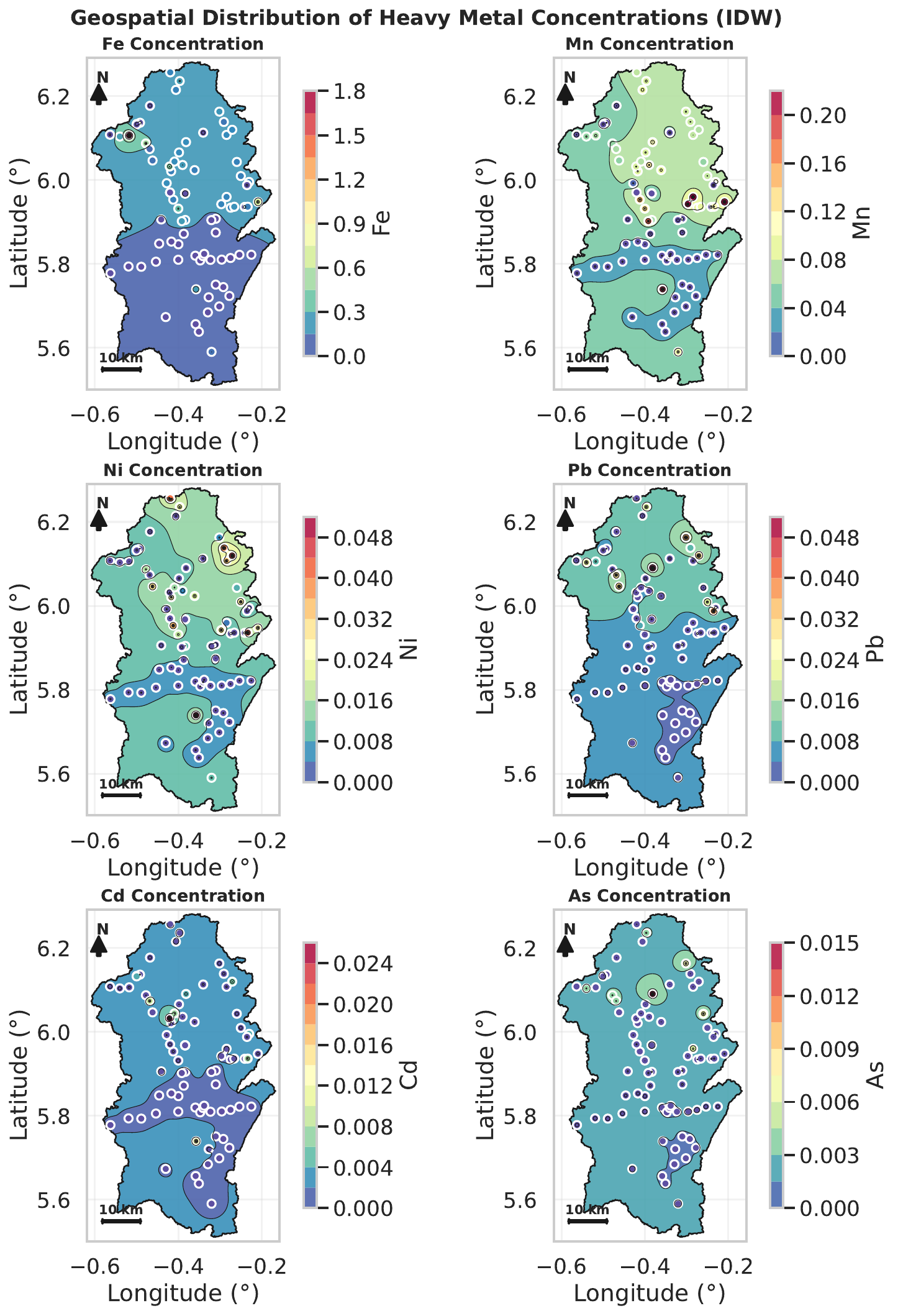}
\caption{Spatial distribution of dissolved heavy metal concentrations across the
Densu Basin derived using IDW interpolation.
Panels show interpolated concentrations of Fe, Mn, Ni, Pb, Cd, and As based on
groundwater samples collected at discrete locations (circles). The maps
highlight pronounced spatial heterogeneity, reflecting the combined influence of
aquifer lithology, groundwater flow pathways, redox conditions, and localised
anthropogenic inputs. Concentration patterns provide a hydrogeochemical context
for subsequent Bayesian health risk integration, but do not, on their own,
represent cumulative exposure or probabilistic health risk.}
\label{fig:metal_maps}
\end{figure*}

\begin{figure*}
\centering
\includegraphics[width=\textwidth]{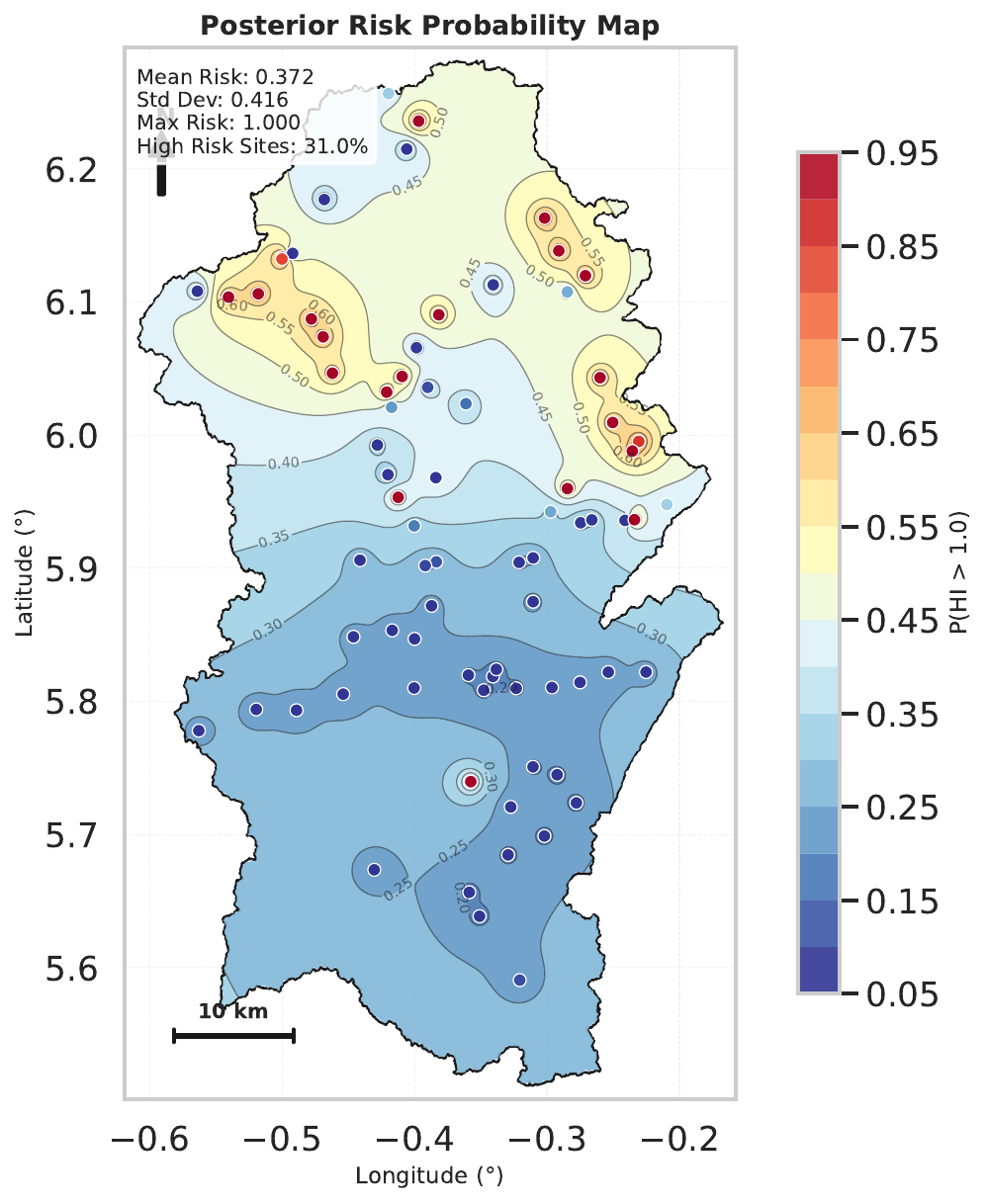}
\caption{Posterior predictive spatial distribution of cumulative non-carcinogenic
health risk for children, expressed as the probability of Health Index
exceedance, $P(\mathrm{HI}_{\mathrm{children}} > 1)$. Site-level posterior risk
probabilities, computed from the Bayesian health risk model
(Eq.~\eqref{eq:site_risk_probability}), were interpolated using inverse distance
weighting to produce a continuous risk surface constrained to the basin
boundary. Contours denote equal-risk probability levels, while point symbols
indicate sampling locations coloured by their posterior risk estimates. Summary
statistics reported on the map (mean, standard deviation, maximum risk, and
proportion of high-risk sites) quantify spatial variability and uncertainty in
cumulative exposure across the basin.}
\label{fig:posterior_risk_map}
\end{figure*}

Figure~\ref{fig:metal_maps} presents the spatial distribution of dissolved heavy
metal concentrations across the Densu Basin, interpolated using the IDW method, while Fig.~\ref{fig:posterior_risk_map} shows the corresponding posterior predictive probability of cumulative health risk exceedance, $P(\mathrm{HI}_{\mathrm{children}} > 1)$.
Together, these maps provide complementary perspectives: the former characterises
the hydrogeochemical structure of groundwater contamination, whereas the latter integrates contaminant co-occurrence, toxicological weighting, and uncertainty within a Bayesian framework to yield a spatially explicit risk metric.

The individual metal concentration maps (Fig.~\ref{fig:metal_maps}) reveal marked
spatial heterogeneity across the basin, reflecting the combined influence of
lithological controls, groundwater flow pathways, and localised anthropogenic
inputs. Elevated concentrations of Fe and Mn are observed primarily in the
central and northern sectors, consistent with reducing conditions commonly
associated with weathered crystalline basement aquifers and prolonged residence
times \citep{Edmunds2012, appelo2004geochemistry}. 
In contrast, Pb and As
display more spatially clustered anomalies, particularly in the north-eastern
and north-western portions of the basin. Such patterns are indicative of mixed
geogenic and anthropogenic influences, including mineralised lithologies,
legacy contamination, and interactions with surface activities along recharge
zones.

Nickel and cadmium exhibit comparatively localised hotspots with broader regions
of low background concentration. This spatial fragmentation suggests point or
line sources rather than basin-wide mobilisation processes, potentially linked
to agricultural inputs, waste disposal, or small-scale industrial activities. Importantly, while several metals show elevated concentrations in isolation, the maps also highlight areas where multiple
contaminants spatially overlap, underscoring the limitation of single-contaminant interpretations in assessing cumulative exposure.

The posterior risk probability map (Fig.~\ref{fig:posterior_risk_map}) synthesises
these heterogeneous concentration fields into a single probabilistic indicator
of health concern. Unlike the individual metal maps, which represent measured
concentrations at discrete locations, the posterior risk surface reflects the
joint posterior predictive distribution defined in
Eq.~\eqref{eq:site_risk_probability} and interpolated according to
Eq.~\eqref{eq:idw_risk}. Areas of elevated posterior risk are concentrated
predominantly in the northern and north-central parts of the basin, where the
probability of $\mathrm{HI} > 1$ locally exceeds $0.5$, indicating a greater than
even posterior likelihood of non-carcinogenic risk.

The spatial coherence of these higher-risk zones is notable. Regions exhibiting
moderate-to-high posterior risk coincide with areas where Pb, As, and Cd
concentrations overlap spatially, even when individual concentrations do not
uniformly exceed guideline values. This outcome illustrates a central advantage
of the Bayesian framework: risk emerges not solely from extreme values of a
single contaminant, but from the combined effect of multiple metals under
uncertainty. The posterior mean risk of 0.372 and standard deviation of 0.416,
reported on Fig.~\ref{fig:posterior_risk_map}, further indicate substantial spatial
and inferential variability, reflecting both heterogeneity in groundwater
chemistry and uncertainty propagated from the Bayesian model.

From a hydrogeological perspective, the north–south gradient is evident in the risk
surface aligns with known variations in aquifer characteristics across the Densu
Basin. Northern sectors are characterised by relatively shallow groundwater
tables, fractured bedrock aquifers, and stronger surface–groundwater
interactions, conditions that can enhance both contaminant mobilisation and
human exposure \citep{macdonald2012quantitative}. In contrast, the southern basin
exhibits generally lower posterior risk probabilities, consistent with shorter
residence times, greater dilution, and reduced accumulation of trace metals
along flow paths.

It is also noteworthy that approximately 31\% of sampling locations are
classified as high-risk sites based on posterior probabilities exceeding 0.5.
This proportion does not imply uniform hazard across these areas but instead
quantifies the degree of epistemic uncertainty associated with cumulative
exposure. Such probabilistic characterisation provides a more informative basis for risk management than binary exceedance mapping, allowing decision-makers to prioritise areas where both risk magnitude and uncertainty are high
\citep{helton1996treatment, Ascough2008}.

Overall, the juxtaposition of individual metal concentration maps with the
posterior predictive risk surface demonstrates how Bayesian integration alters
the spatial narrative of groundwater quality. While raw concentration maps are
essential for identifying contaminant sources and hydrogeochemical processes,
the posterior risk map offers a policy-relevant synthesis that explicitly
accounts for co-exposure and uncertainty. This distinction is particularly
important in data-sparse settings such as the Densu Basin, where conventional
deterministic mapping may underestimate cumulative health risks or obscure
priority intervention zones.

\subsection{Implications for Water Quality Management and Public Health Policy}

Beyond statistical inference, the Bayesian results have direct implications for
groundwater management and public health policy within the Densu Basin. The
posterior risk attribution indicates that Pb, As, and Cd consistently dominate
the cumulative non-carcinogenic risk for children, suggesting that mitigation
efforts should prioritise contaminants that combine elevated toxicity with a
high posterior probability of contribution to overall exposure. This finding
highlights the limitations of regulatory approaches based solely on
concentration exceedances of individual contaminants, which may overlook
co-exposure effects and associated uncertainty.

From a policy perspective, the posterior probability of Health Index exceedance,
$P(\mathrm{HI}_{\mathrm{children}} > 1)$, provides a more informative decision
metric than deterministic thresholds alone. Rather than classifying sites as
simply compliant or non-compliant, probabilistic exceedance estimates enable
risk-based prioritisation, allowing regulators to distinguish between locations
where exceedance is highly likely, marginal, or uncertain. Such differentiation
is particularly valuable in settings where monitoring density is limited and
management resources are constrained.

The probabilistic ranking of metals further suggests that targeted monitoring and
intervention strategies should focus on Pb- and As-enriched groundwater zones,
especially in areas influenced by urban expansion, informal mining activities,
or hydrogeochemical conditions conducive to metal mobilisation, such as reductive
dissolution of iron-bearing minerals. In these zones, risk-informed measures
including source protection, wellhead treatment, selective well replacement, or
point-of-use filtration may be prioritised over basin-wide interventions that
assume spatially uniform risk.

More broadly, the Bayesian framework adopted in this study supports adaptive water
quality management by explicitly incorporating uncertainty and co-exposure into
risk assessment. Posterior uncertainty estimates can be used to guide the design
of monitoring programmes, for example by identifying locations where additional
sampling would most reduce uncertainty in risk classification. This adaptive,
evidence-based approach offers particular value in data-limited regions, where
incremental improvements in information must be balanced against logistical and
financial constraints. By moving beyond binary regulatory thresholds, the
framework provides a transparent basis for staged and proportionate public
health interventions.

\section{Conclusions}
\label{sec:conclusion}

This study applied a Bayesian probabilistic framework to assess
non-carcinogenic health risks associated with multi-metal groundwater
exposure among children in the Densu River Basin, Ghana. By integrating
groundwater chemistry, the child Health Index, posterior inference, and
spatial risk mapping, the analysis provides a probabilistic
characterisation of potential health risk under the specified exposure
and modelling assumptions. The findings identify Pb, As, and Cd as the
principal contributors to the modelled cumulative Health Index and
indicate that the likelihood of exceeding the conventional health-risk
threshold varies across sampling locations. These results demonstrate
the importance of considering contaminant co-occurrence and uncertainty
when interpreting groundwater-related health risks, rather than relying
exclusively on individual metal concentrations or deterministic
threshold classifications.

The study also illustrates how site-specific posterior risk estimates
can be represented spatially to support the identification of areas
that may warrant further investigation. The resulting maps provide a
basin-scale view of the distribution of estimated risk, while their
spatial detail remains constrained by the number and distribution of
groundwater sampling locations. The interpolated surfaces should
therefore be interpreted alongside the observed site-level estimates,
rather than as independently validated predictions of risk at
unsampled locations.

Several limitations define the scope of these conclusions. Groundwater
samples were collected during a single sampling campaign in January
2020, corresponding to the dry season. Consequently, the analysis does
not capture seasonal variation in groundwater metal concentrations
that may arise from changes in recharge, dilution, groundwater levels,
and geochemical conditions. The reported risk estimates therefore
characterise the sampled conditions and cannot establish whether the
same risk patterns persist throughout the year. In addition, the
assessment is based on the specified Health Index formulation and
regression structure, which do not explicitly represent non-linear
exposure--risk relationships, interactions among contaminants, or
exposure pathways beyond those included in the assessment.

Future work should incorporate repeated wet- and dry-season sampling
to establish the temporal stability of contaminant concentrations and
associated risk estimates. Further development of spatial Bayesian
models could also provide a more explicit representation of spatial
dependence and uncertainty between sampling locations. Incorporating
additional exposure pathways and evaluating alternative exposure--risk
formulations would further strengthen the assessment.

Overall, this study provides a probabilistic assessment of potential
child health risks associated with multi-metal groundwater exposure
in the Densu River Basin. Its principal contribution is the
characterisation of uncertainty in Health Index exceedance and the
identification of contaminant priorities and locations for further
investigation. These outputs can inform groundwater monitoring and
risk-management decisions, provided that the estimates are interpreted
in the context of the study's temporal coverage, sampling distribution,
and modelling assumptions.

\section{Statements and Declarations}
\subsection*{Competing interests}
The authors declare that they have no competing financial interests or personal relationships that could have influenced the work reported in this study.

\subsection*{Funding}
The authors declare that no funds, grants, or other support were received during the preparation of this manuscript.

\subsection*{Author Contributions}
\textbf{T. Ansah-Narh} conceived the study, designed the methodology, performed the Bayesian modelling and spatial analysis, interpreted the results, and led the writing of the manuscript.
\textbf{G. Y. Afrifa} contributed to data acquisition, preprocessing, and the conventional health risk assessment framework.
\textbf{J. B. Tandoh} assisted with hydrogeological interpretation, groundwater chemistry analysis, and contextualisation of results within the study area.
\textbf{E. N. N. Nortey} contributed to statistical analysis, validation of model outputs, and critical review of the manuscript.
\textbf{D. Adjei} contributed to the interpretation of the results and critically reviewed the manuscript for intellectual content.
\textbf{R. Opoku-Sarkordie} supported the spatial analysis and mapping components and assisted with data visualisation.
\textbf{R. Davis} provided conceptual guidance, contributed to the interpretation of results, and critically reviewed the manuscript for intellectual content.
 
\subsection*{Acknowledgements}

T.A-N. acknowledges the Ghana Space Science and Technology Institute (GSSTI) for providing access to its High-Performance Computing (HPC) cluster, which supported the computational analysis. The authors also acknowledge the French Embassy in Ghana for its support through the Artificial Intelligence for Sustainable Development (AI4SD) initiative, which promotes the application of artificial intelligence to sustainable development challenges.

\section*{Data availability}
The datasets generated and/or analysed during the current study are not publicly available, as the project is ongoing, but are available from the corresponding author upon reasonable request.

\end{document}